\documentclass[manuscript]{aastex}
\usepackage{graphics}      
\usepackage{graphicx}      
\usepackage{longtable}     
\usepackage{url}           
\usepackage{bm}            
\usepackage[usenames]{color}
\usepackage[OT2,T1]{fontenc}
\usepackage{amssymb}
\usepackage{wasysym}    
\usepackage{ccaption}
\usepackage{float}
\usepackage{multirow}
\usepackage{natbib}

\newcommand{\xone}{\objectname{LMC~X--1}}
\newcommand{\xthree}{\objectname{LMC~X--3}}
\newcommand{\cyg}{\objectname{Cyg~X--1}}

\definecolor{plum}{rgb}{0.5,0.1,0.5}
\definecolor{brightpurple}{rgb}{0.7,0.2,0.7}
\definecolor{orange}{rgb}{1,0.25,0}
\definecolor{green}{rgb}{0.3,0.7,0.2}

\newcommand{\response}[1]{#1}

\renewcommand{\sun}{\odot}

\begin{document}

\title{The nature and cause of spectral variability in LMC~X--1}
\author{L. Ruhlen}
\affil{Astronomy Department, UC Santa Cruz, Santa Cruz, CA 95064}
\email{lruhlen@ucsc.edu}

\author{D. M. Smith\altaffilmark{2}}
\affil{Physics Department, UC Santa Cruz, Santa Cruz, CA 95064}
\email{dsmith@scipp.ucsc.edu}

\author{J. H. Swank\altaffilmark{3}}
\affil{Goddard Space Flight Center, NASA, Astrophysics Science Division, Greenbelt MD 20771, USA }
\email{swank@lheavx.gsfc.nasa.gov}

\begin{singlespace}
\begin{abstract}
We present the results of a  long-term observation campaign of the
extragalactic wind-accreting black-hole X-ray binary LMC~X--1, using the
Proportional Counter Array on the Rossi X-Ray Timing Explorer
(RXTE). The observations show that LMC~X--1's 
accretion disk exhibits an anomalous temperature-luminosity relation. 
We use deep archival RXTE observations to show that large movements
across the temperature-luminosity space occupied by the system can
take place on time scales as short as half an hour. 
These changes cannot be adequately explained by perturbations that
propagate from the outer disk on a viscous timescale. We propose
instead that the apparent 
disk variations reflect rapid fluctuations within the Compton
up-scattering coronal material, which occults the inner parts of the disk. 
The expected relationship between the observed disk luminosity and apparent
disk temperature derived from the variable occultation model is
quantitatively shown to be in good agreement with  
the observations. Two other observations support this
picture: an inverse correlation between the flux in the power-law
spectral component and the fitted inner disk temperature, and a
near-constant total photon flux, suggesting that the inner disk is not
ejected when a lower temperature is observed. 

\end{abstract}


\section{Introduction}

LMC~X--1 is (with \cyg) one of only two known persistently luminous x-ray binaries 
consisting of a black hole accreting the wind of a massive blue star.
The $10.91 \pm 1.41 M_{\sun}$ black hole is in a $3.90917 \pm 0.00005$ day
orbit \citep{lmcx1-orosz-2009} about an O7~III companion
\citep{lmcx1-cowley-1995}. As the companion both drives a strong wind
and is far from filling its Roche lobe, wind accretion feeds the black
hole \citep{lmcx1-and-lmcx3-nowak-2000}.  The system, which accretes at an average of
$0.16 L_{Edd}$ \citep{lmcx1-gou-2009} has never been observed in the
low/hard state \citep{lmcx1-and-lmcx3-wilms-2001}.  \citet{lmcx1-gou-2009} 
used x-ray data, including some of the {\it RXTE} observations we use here,
to derive a spin parameter $\alpha = 0.92 (+0.05, -0.07)$ for the black hole
in LMC~X--1 using fits to the disk blackbody component of the spectrum.

\xone's persistent occupation of the high/soft state 
contrasts with the behavior of \cyg, which it closely resembles 
in other respects. Cyg~X--1 harbors a $\sim 10M_{\odot}$ black hole 
\citep{herrero-1995} in a $5.6$ day orbit
\citep{cygx1-paczynski-1974}  
about an O9.7 Iab companion \citep{cygx1-bolton-1972}.  
Though \cyg\ is
also a wind-accreting system, its typical luminosity in the high state
 is $\sim 0.03 L_{Edd}$ \citep{cygx1-gierlinski-1999}, which is
much lower than \xone's.  
This implies that the black hole in LMC~X--1 accretes at a higher rate
than its counterpart in \cyg, perhaps because
the companions' wind speeds  and densities differ, or that 
accretion mechanisms with different radiative efficiencies are taking place.
In addition, \cyg\ exhibits well-documented transitions between  
the high/soft state and the low/hard state \citep{cygx1-agrawal-1972,
  cygx1-baity-1973, cygx1-heise-1975, cygx1-holt-1976,
  cygx1-holt-1975, cygx1-sanford-1975}.
LMC~X--1, in contrast, displays persistent soft-state emission
\citep{lmcx1-and-lmcx3-nowak-2000,lmcx1-and-lmcx3-wilms-2001}. 
While both \xone\ and \cyg's softest states can be fitted with a power
law plus disk blackbody model, they also differ markedly.  The
traditional soft state model, in which the disk blackbody dominates
the energy spectrum, accurately 
fits \xone's spectra \citep{ebisawa-1989, yao-2005}. \cyg's
spectra, however, are 
energetically dominated by the power law in both the hard and soft states.
Thus, while these two systems have similar companions and orbital
properties, they differ in their spectral characteristics and evolution.

An even more interesting comparison can be made between \xone\ and \xthree,
a black-hole binary that accretes via Roche-lobe overflow at comparable
luminosity, as a means of 
highlighting the observational differences between disk and wind accretion.
Much of this paper will be devoted to pointing out and interpreting these
differences.
  
\xthree\  is a persistently bright x-ray binary in the Large
Magellanic Cloud, and is the only other system with a dynamically
confirmed black hole.  As such, it provides a useful counterpoint to the 
wind-accreting black hole binary systems mentioned above.  
LMC~X--3 is a $9 M_{\odot}$ black hole \citep{lmcx3-vanderKlis-1985}
with an evolved B5 companion \citep{lmcx3-soria-2001}.  A 
detailed listing of the system's dynamical properties, along with
those of \xone\ and \cyg, is provided in Table \ref{tb:all-systems-orbital-params}. 
Differences between the systems are expected to reflect
differences in the accretion flows resulting from wind
versus Roche-lobe overflow accretion: LMC~X--3 accretes high angular
momentum material through its first     
Lagrange point, while LMC~X--1 accretes 
low angular momentum material from its companion's stellar wind.
LMC~X--3 and LMC~X--1 share spectral 
signatures of a persistent accretion disk, though 
 LMC~X--1's disk spectrum defies the $L
\propto T^4$ relation \citep{diskbb2, dunn-2010}.

\begin{table}
  \centering
   \begin{minipage}{0.99\textwidth}
    \caption{System parameters for LMC~X--1, LMC~X--3, and
    Cygnus~X--1, with references in parenthesis. The Eddington
    luminosity for each source was calculated from black hole mass.}  
      \begin{tabular}{|p{20mm}|r|r|r|}
	\hline
        & \textbf{\xone} & \textbf{\xthree} & \textbf{\cyg}\\
	\cline{1-4}
	\hline
	Distance (kpc)&$48.10 \pm 2.22$ (1)
        & $52.0 \pm 0.6$ (2) & $2.5$ (3)\\
	\hline
	Black Hole Mass ($M_{\sun}$)&$10.91 \pm 1.41
        $ (1)& $11.1 \pm 1.4$ (4)  &$10.1$ (5)\\
	\hline
	System Inclination (degrees)& $36.38 \pm
        1.92^{\circ}$ (1)& $50.0^{\circ} \leq i \leq 70.0^{\circ}$
        (12) & $36^{\circ} < i < 67^{\circ}$ (11)\\
	\hline
	Companion Type& O7 III (8) & B5 IV
        (6)& O9.7 Iab (5) \\
	\hline
	Companion Mass ($M_{\sun}$)& $31.79 \pm 3.48$
        (1) & $4.0 \leq M \leq 4.7$ (6) & $17.8$
         (5)\\
	\hline
	Companion Radius ($R_{\sun}$)& $17.0 \pm 0.8$ (1)  & $4.4$
        (6)& $22.7 \pm 2.3$ (7) \\
	\hline
	Orbital Period (days)&$3.90917 \pm
        0.00005$ (1)
        &$1.70479 \pm 0.00004$ (5)  &
        $5.6$ (3) \\
	\hline
	Average Soft State Luminosity (\% L$_{Edd}$) &$\sim 16\%$ (8) &$\leq 30 \%$
	(9)  & $\sim 3\%$ (10)\\
	\hline
    \end{tabular} 

      \label{tb:all-systems-orbital-params}

      (1) \citet{lmcx1-orosz-2009};
      (2) \citet{dibenedetto-1997};
      (3) \citet{cygx1-paczynski-1974};
      (4) \citet{gierlinski-2001};
      (5) \citet{herrero-1995};
      (6) \citet{lmcx3-soria-2001};
      (7) \citet{cygx1-companion-radius};
      (8) \citet{lmcx1-gou-2009};
      (9) \citet{lmcx1-and-lmcx3-nowak-2000};
      (10) \citet{cygx1-gierlinski-1999};
      (11) \citet{davis-1983};
      (12) \citet{lmcx3-vanderKlis-1985}.

  \end{minipage}
  \end{table}

This paper examines how \xone's
spectral properties and evolution fit with, and possibly extend,
existing black hole binary accretion models.  
In particular, our results ultimately indicate
that \xone's spectral behavior is consistent with sporadic
obscuration of the innermost part of a stable accretion disk.

Section 2 of this paper presents the data processing and spectral
fitting techniques 
employed in obtaining the results presented in Section 3.  Section 4
discusses how \xone's unique spectral properties can be reconciled
with existing accretion flow models, while Section 5 presents our
conclusions and possible avenues for further investigation.

\section{Observations and Data Analysis}
The data analyzed in this work fall into two categories.  The bulk of
the observations come from our twice-weekly RXTE monitoring campaign of
\xone\ and \xthree.  The \xone\ data were complemented by long,
individual RXTE observations available on the HEASARC.  These data
were reduced and analyzed using identical processes.  The two
classes of \xone\ observations prove useful for probing different
timescales of spectral variations.

\subsection{Observations}
The main observing campaign consisted of brief, uninterrupted
  observations of \xone\ and \xthree, conducted twice a week since August  
  2007 and March 2006, respectively.  These observations 
  ranged in length between 
  506 and 5470 seconds, with average lengths of 1661
  seconds for \xone\ and 1877 seconds for \xthree.   
The \xone\ observations were offset from the source
by 15.764' to reduce contamination from the nearby
pulsar PSR B0540-69.  
These twice-weekly observations give reliable pictures of the typical
 X-ray spectra of \xone\ and \xthree\ over significant periods of time.
While this is a major strength of these data sets, it also a weakness,
in that they only have the capacity to reveal spectral variations that
emerge over a minimum of several days.  

To examine shorter timescale
variations in \xone, we take advantage of the seventeen deep
archival RXTE observations listed in Table \ref{tb:archival-observations}. 
We ``chopped'' these data sets into shorter spectra (less than 90~min)
corresponding to one orbit of the RXTE spacecraft.
These chopped archival
observations can reveal spectral changes that emerge over fractions of
a day.

\begin{center}
  \begin{table}[h]
      \hfill{}
      \begin{tabular}{|p{30mm}|p{30mm}|p{30mm}|p{30mm}|p{30mm}|}
	\hline 
	\textbf{Observation ID} & \textbf{UT (yyyy-mm-dd)}
	&\textbf{Start Date (MJD)} & \textbf{Total Exposure (s)} &
	\textbf{Sub-Observations} \\ 
	\hline
	20188-01-02-00  & 1996-12-30  & 50447.438  & 9919  & 4 \\
	\hline
	20188-01-03-00  & 1997-01-18  & 50466.347  & 9941  & 3 \\
	\hline
	20188-01-05-00  & 1997-03-09  & 50516.354  & 10189 & 4 \\
	\hline
	20188-01-06-00  & 1997-03-21  & 50528.004  & 10625 & 4  \\
	\hline
	20188-01-07-00  & 1997-04-16  & 50554.193  & 11484 & 3  \\
	\hline
	20188-01-14-00  & 1997-09-09  & 50700.695  & 10174 & 4  \\
	\hline
	20188-01-18-00  & 1997-10-10  & 50731.611  & 11436 & 3  \\
	\hline
	30087-01-02-00  & 1998-01-25  & 50838.667  & 10017 & 3  \\
	\hline
	30087-01-03-00  & 1998-02-20  & 50864.370  & 9948  & 4 \\
	\hline
	30087-01-04-00  & 1998-03-12  & 50884.495  & 10007 & 7  \\
	\hline
	30087-01-06-00  & 1998-05-06  & 50939.149  & 11396 & 4  \\
	\hline
	30087-01-07-00  & 1998-05-28  & 50961.101  & 5966  & 2 \\
	\hline
	30087-01-08-00  & 1998-06-28  & 50992.097  & 9921  & 4 \\
	\hline
	30087-01-09-00  & 1998-07-19  & 51013.909  & 10094 & 4  \\
	\hline
	80118-01-06-02  & 2004-01-11  & 53015.340  & 10908 & 3  \\
	\hline
	80118-01-07-00  & 2004-01-12  & 53016.258  & 10788 & 3  \\
	\hline
	80118-01-08-00  & 2004-01-10  & 53014.236  & 6608  & 2 \\
     \hline
      \end{tabular}
      
	\hfill{}
	\caption{Summary of the deep archival RXTE observations used
	to constrain the power law index, and to find upper limits on
	the apparent timescale of disk luminosity evolution. The
	former used only the parts of the spectra above 12 keV, while
	the latter relied only on regions of the spectra below 12 keV.}
	\label{tb:archival-observations}    
  \end{table}
\end{center}

\subsection{Data Analysis}

All data were reduced using standard HEASOFT v6.7 tools.  Data taken within
30 minutes after the satellite exited the South Atlantic Anomaly
(SAA), or when the source was at less
than 10$^{\circ}$ elevation above Earth's limb, were excluded from subsequent
analysis. All observations only considered data from the second
proportional counter unit (PCU2) of RXTE's Proportional Counter Array
(PCA), due to its consistent performance throughout the mission.
The mission-long faint background model
\texttt{pca\_bkgd\_cmfaintl7\_eMv20051128.mdl}  
was used in this analysis, as 
the total count rate for both LMC sources lies below 40
photons/s/PCU. Data were extracted using the \texttt{Standard2}
data mode.   

\begin{figure}
  \includegraphics[width=0.8\textwidth,angle=0]{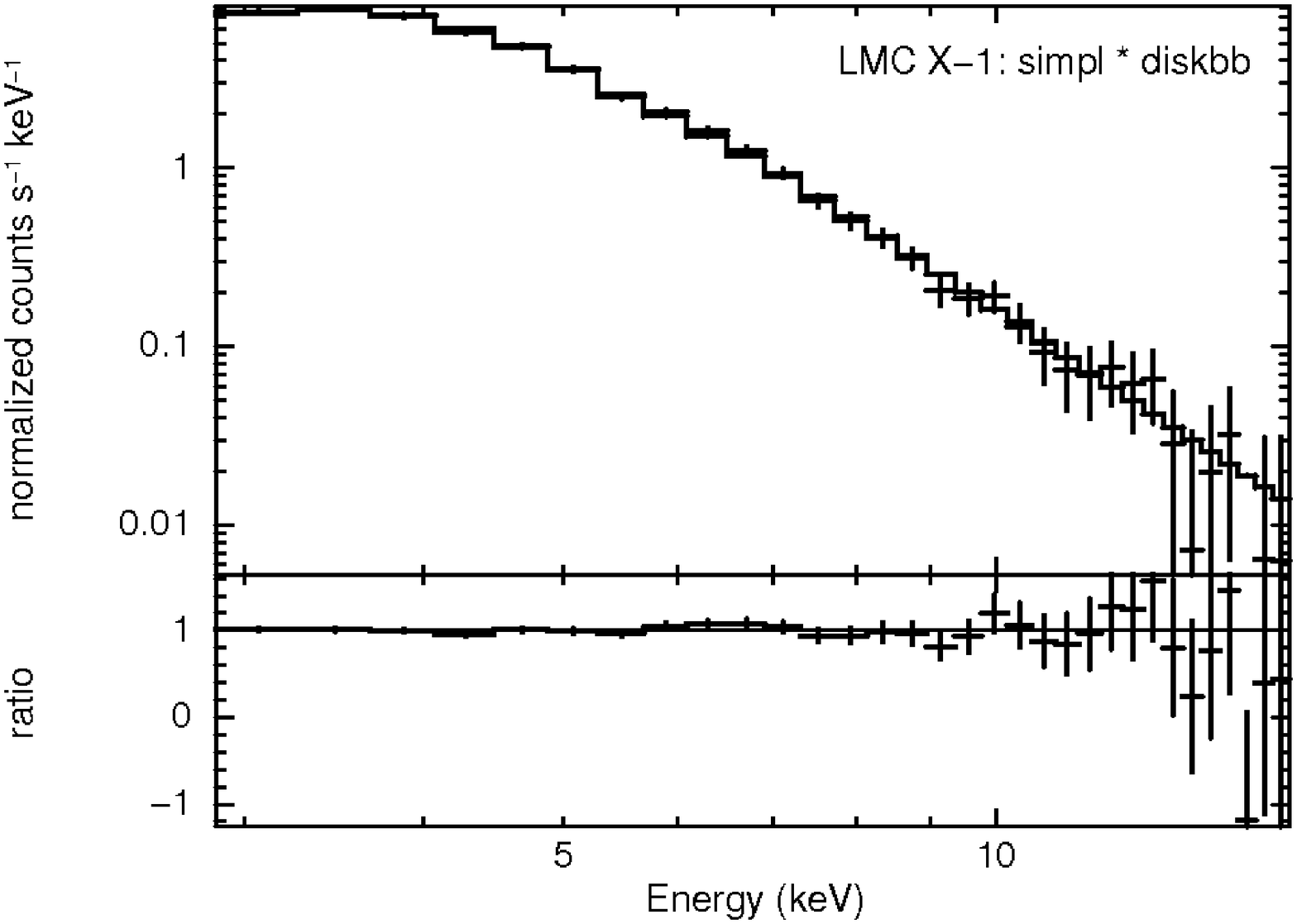}
  \captionstyle{normal}
  \caption{Fit and data-to-model ratio for \protect\xone\ data fit by the multi-colored disk and \protect\texttt{simpl} (power-law) spectral model.}
  \label{fig:lmcx1-fit-with-simpl-and-diskbb}
\end{figure}

\begin{figure}
  \includegraphics[width=0.8\textwidth,angle=0]{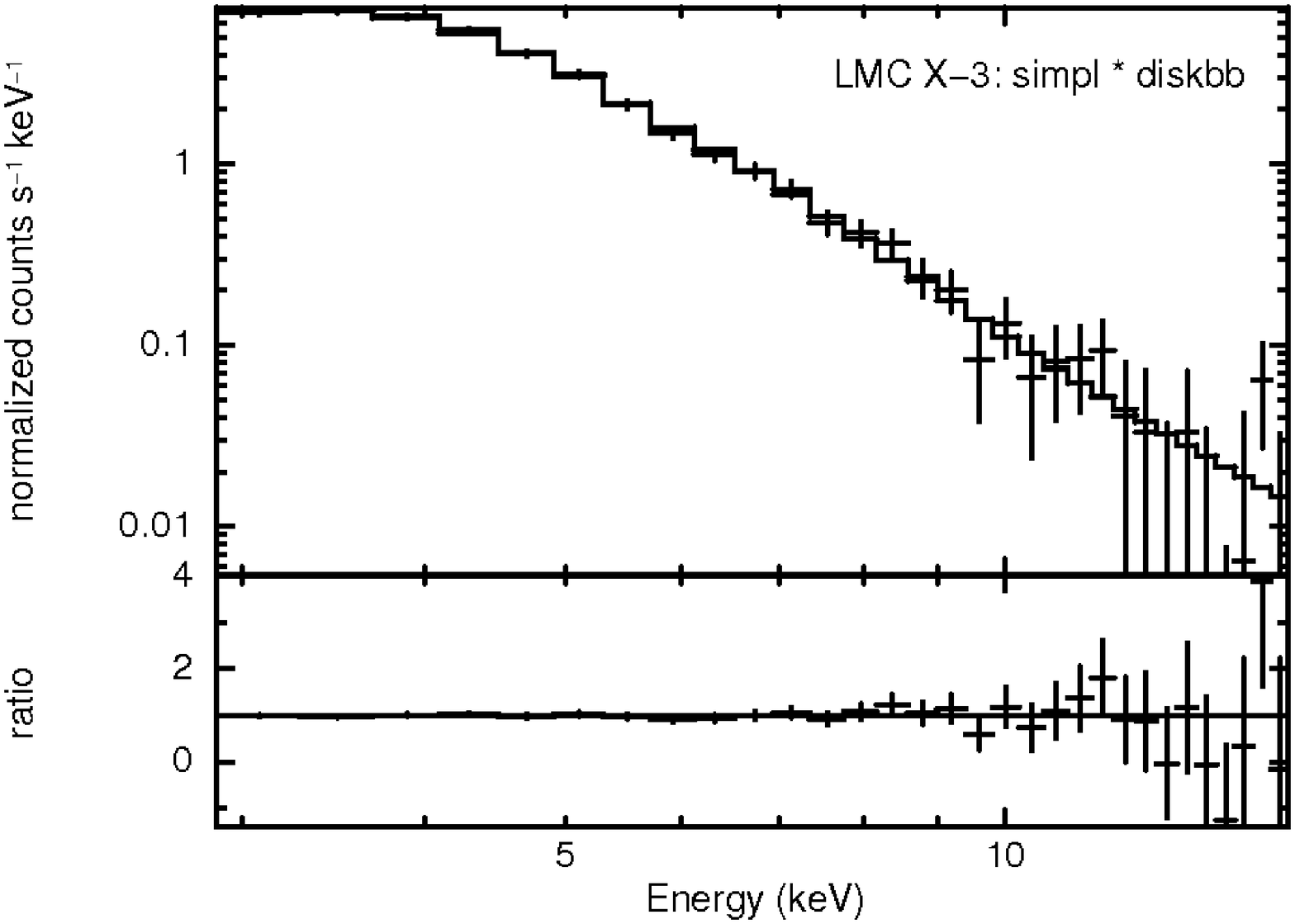}
  \caption{Fit and data-to-model ratio for \protect\xthree\ data fitted by
  multi-colored disk and \protect\texttt{simpl} (power-law) 
  spectral model.\label{fig:lmcx3-diskbb-and-simpl-fit}} 
\end{figure}

We used uniform XSPEC v12.5.1 \citep{Arnaud1996}
fitting procedures for all data.  
Per RXTE recommendation \citep{Jahoda2006}, systematic errors
of 1\% were added in quadrature to the Poisson counting noise errors.
Because the detector response is poorly understood within the
first three PCA channels, we exclude those channels from our analysis,
resulting in a lower energy threshold just under 3 keV.
Above 16 keV, the signal to noise ratio for both sources is too low
to provide useful fits to the data, so our fits also exclude
all data for energies over 16 keV.  
A 29\% decrease in the PCA
efficiency \citep{Jahoda2006}
relative to an on-axis source resulted from the 15.764'
off-axis pointing of our \xone\ observations.
All fluxes derived from the off-axis \xone
observations were multiplied by a factor of 1.4 to account for this
efficiency factor.    
Representative spectra, model fits, and fit residuals for \xone\ and
\xthree\ are presented in Figures
\ref{fig:lmcx1-fit-with-simpl-and-diskbb} and 
\ref{fig:lmcx3-diskbb-and-simpl-fit}, respectively.

Spectra were fitted with a disk blackbody plus power law emission model. 
The first
spectral component
is a multicolored blackbody spectrum generated by the accretion
disk, which is accounted for using the standard
\texttt{diskbb} XSPEC model \citep{diskbb2}.  
The \texttt{simpl} XSPEC model \citep{simpl} accounts for the power-law
emission, which extends to high energies, and which is thought to be caused by Compton
up-scattering of disk photons through a population of hotter
electrons \citep{shapiro-2-comp-flow}. A pure power-law
model runs the risk of 
promoting spectral fits that inaccurately trade disk luminosity for
coronal energy flux.
The XSPEC \texttt{simpl} model avoids this by turning over at low energies, 
and by convolving an arbitrary input photon spectrum with the
specified Compton scattering prescription. It is, therefore, less
prone to overestimating the amount of 
non-thermal emission. \texttt{Simpl} can account for either pure photon
up-scattering, or for both up- and down-scattering.  Because the
average energy of coronal electrons is two orders of magnitude higher
than the highest disk blackbody temperature, coronal emission from our
sources can be safely approximated by pure up-scattering. 

Finally, to account for photoelectric absorption 
within the intervening ISM, we convolve the compound model
with \texttt{phabs}. 
\response{Our approach does not conflict with the findings
of \citet{levine-and-corbet-2006} and
\citet{hanke-and-wilms-2010}, which detect orbital phase variations in
the absorption column towards LMC X-1. These orbital phase
variations suggest that LMC X-1's companion 
drives a significant wind.  While the variable column density
 significantly effects LMC X-1's observed soft x-ray emission, it
does not impact the higher energy range examined in this analysis.}

Table \ref{tb:summary-of-fit-params} lists the fitting parameters and
results from the following fitting procedure.  The \texttt{simpl} power law
index, the absorption column, and the inverse Compton scattering flag
are all frozen at the listed values. The absorption column densities
are drawn from \citet{lmcx1-and-lmcx3-nowak-2000} for both \xone\ and
\xthree.  The innermost disk temperature 
starts, but is not frozen, at the physically reasonable energy of 1
keV.  The spectral fits output the inverse Compton photon scattering
fraction, the innermost disk temperature, the overall
normalization of the \texttt{diskbb} model component, and $\chi^2$
goodness-of-fit statistics.  Once the model has been fitted to the data, it
is used to calculate both comprehensive and component-specific photon
and energy fluxes. 

\begin{center}
  \begin{table}[ht]
    {\small
      \hfill{}
      \begin{tabular}{|l|l|c|c|}
	\hline
	&  & \textbf{\xone} & \textbf{\xthree} \\
	\hline
	\multirow{2}{*}{Inputs (frozen)} 
	& nH (\textrm{$cm^{-2}$})  &
	\textrm{$7.2 \times 10^{21}$} (1) 
	& \textrm{$3.2 \times 10^{20}$} (1)
	 \\ \cline{2-4} 

	& $\Gamma$ & \textrm{$2.68$} &
	\textrm{$2.34$} (2) \\ \cline{2-4}

	& Scattering Flag & 1 & 1 \\ 
	\hline
	\multirow{4}{*}{Outputs  (average values)}
	& \textrm{$k T_{e}$} (keV) & \textrm{$0.90 \pm 0.06$} &
	\textrm{$1.07 \pm 0.18$}\\ \cline{2-4}
	& \textrm{$f_{sc}$} & \textrm{$0.13 \pm 0.07$} & \textrm{$0.08
	\pm 0.17$} \\ \cline{2-4}
	& \textrm{$\chi^2/$d.o.f.} & \textrm{$0.79 \pm 0.24$}
	&\textrm{$ 0.79 \pm 0.77$} \\ 
	\hline
    \end{tabular}}
    \hfill{}
    \caption{Input parameters and results for the \protect\texttt{phabs $\times$ (simpl
    $\times$ diskbb)} fits to \protect\xone\ and \protect\xthree\ data, with citations
    in parenthesis.  Uncertainties
    in the outputs correspond to single standard deviations of the
    values about their average. Fixing \protect\texttt{simpl}'s scattering
    flag at a value of 1 forces the fits to account for only
    up-scattering.  (1) \protect\citet{lmcx1-and-lmcx3-nowak-2000}; (2) \protect\citet{smith-2007}} 
    \label{tb:summary-of-fit-params}    
  \end{table}
\end{center}

The low photon counts from both LMC~X--1 and LMC~X--3 limit the usable
energy range in our 1.5 ksec observations to no higher than 16
keV. With counting statistics of this quality, it is not possible to
simultaneously constrain both the power-law index and the inner-disk
temperature.  Since our observations cannot 
 determine all six of the parameters in the
$ \texttt{phabs} \times \texttt{simpl} \times \texttt{diskbb}$
model (see table), we freeze the power-law index, which is the most
stable at long time scales.  
The deep archival observations of LMC~X--1, prior to being chopped into
orbit-by-orbit sections, have sufficient photons to simultaneously
constrain all disk and power-law fit parameters. We use the average
power law index derived from these long observations, which is
consistent with a single value, to fix the
power law index for all of the single-orbit LMC~X--1 spectra.  For
LMC~X--3, we employ the average power law index found in 
the same way by \citet{smith-2007}.  
In both cases, the values for all deep pointings are statistically
consistent with the average.

\section{Results} 

We find that \xone\ remained in the soft state over the entire
observation program, and has an anomalous
disk-temperature-versus-luminosity relation relative to the expectation
of a disk blackbody.

\begin{figure}
  \includegraphics[width=0.7\textwidth]{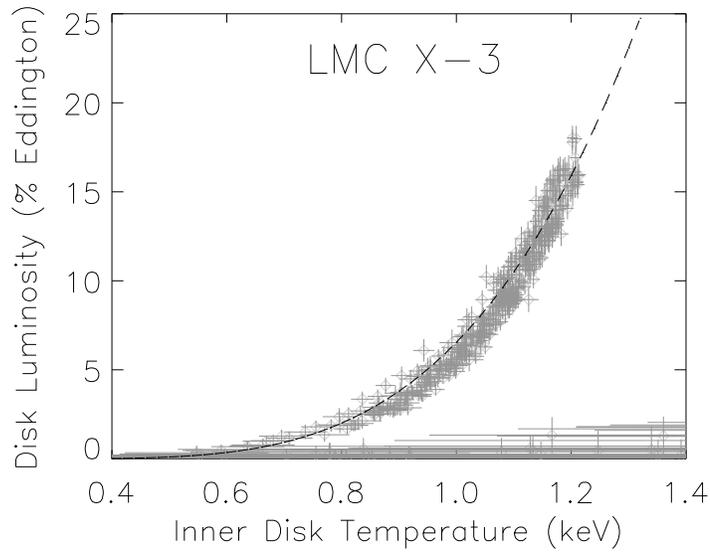}
  \caption{The disk temperature-luminosity relationship observed in
  \protect\xthree. The dashed line is the modified accretion disk
  Stefan-Boltzmann relationship. The points at low
  luminosities and high temperatures, which lie far from the model,
  are due to a transition to the 
  low/hard state towards the end of the observing
  campaign. \label{fig:lmcx3-disktemp-vs-lum-for-flash}}
\end{figure}

\begin{figure}
  \includegraphics[width=0.7\textwidth]{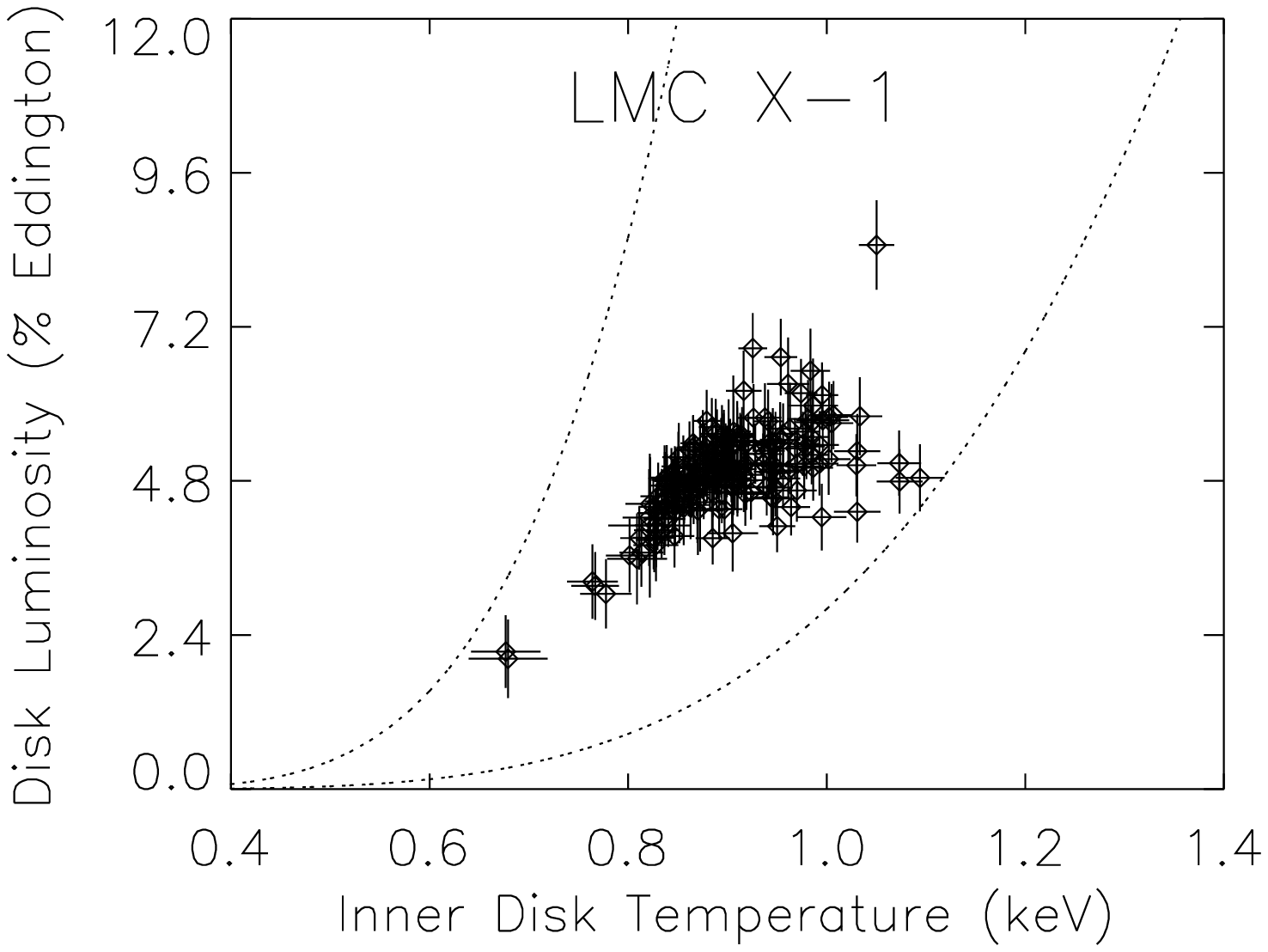}
  \caption{The disk temperature-luminosity relationship observed
    in \protect\xone. The dashed lines are the modified accretion disk
    Stefan-Boltzmann relationship shown in Figure
    \ref{fig:lmcx3-disktemp-vs-lum-for-flash}, and differ only in their
    overall normalization constants.  
    \label{fig:lmcx1-disktemp-vs-lum-for-flash}}
\end{figure}

Figure \ref{fig:lmcx3-disktemp-vs-lum-for-flash} demonstrates how the
luminosity of a typical accretion disk, such as the one present in  
\xthree's soft state, follows the expected modified $L \propto T_{in}^4$
Stefan-Boltzmann relation \citep{diskbb2}. 
The line in Figure \ref{fig:lmcx3-disktemp-vs-lum-for-flash},
which represents this relation, was not fitted to the data beyond finding
a suitable normalization 
constant.  The relation nevertheless fits the data very well, aside
from the highly 
uncertain points at low luminosities and high temperatures.  
These points come from observations taken after Feb. 23, 2009 (MJD
54885), when LMC~X--3 transitioned from a disk-dominated high/soft
state to the low/hard state. The fitting routine we
employ is not appropriate for the low/hard state, during which disk
emission is suppressed and the power-law index hardens significantly.
The poorly constrained values that appear in the lower right corner of
Figure \ref{fig:lmcx1-disktemp-vs-lum-for-flash}, well away from the
Stefan-Boltzmann relation, result from using a high/soft state model
to fit low/hard state data. 
 
Figure \ref{fig:lmcx1-disktemp-vs-lum-for-flash} similarly shows the
disk temperature-luminosity relation for LMC~X--1.  The dashed lines are
two normalizations of the same 
Stefan-Boltzmann relation used in Figure
\ref{fig:lmcx1-disktemp-vs-lum-for-flash}, and clearly  
cannot fit the data. 

\begin{figure}
  \includegraphics[width=0.7\textwidth]{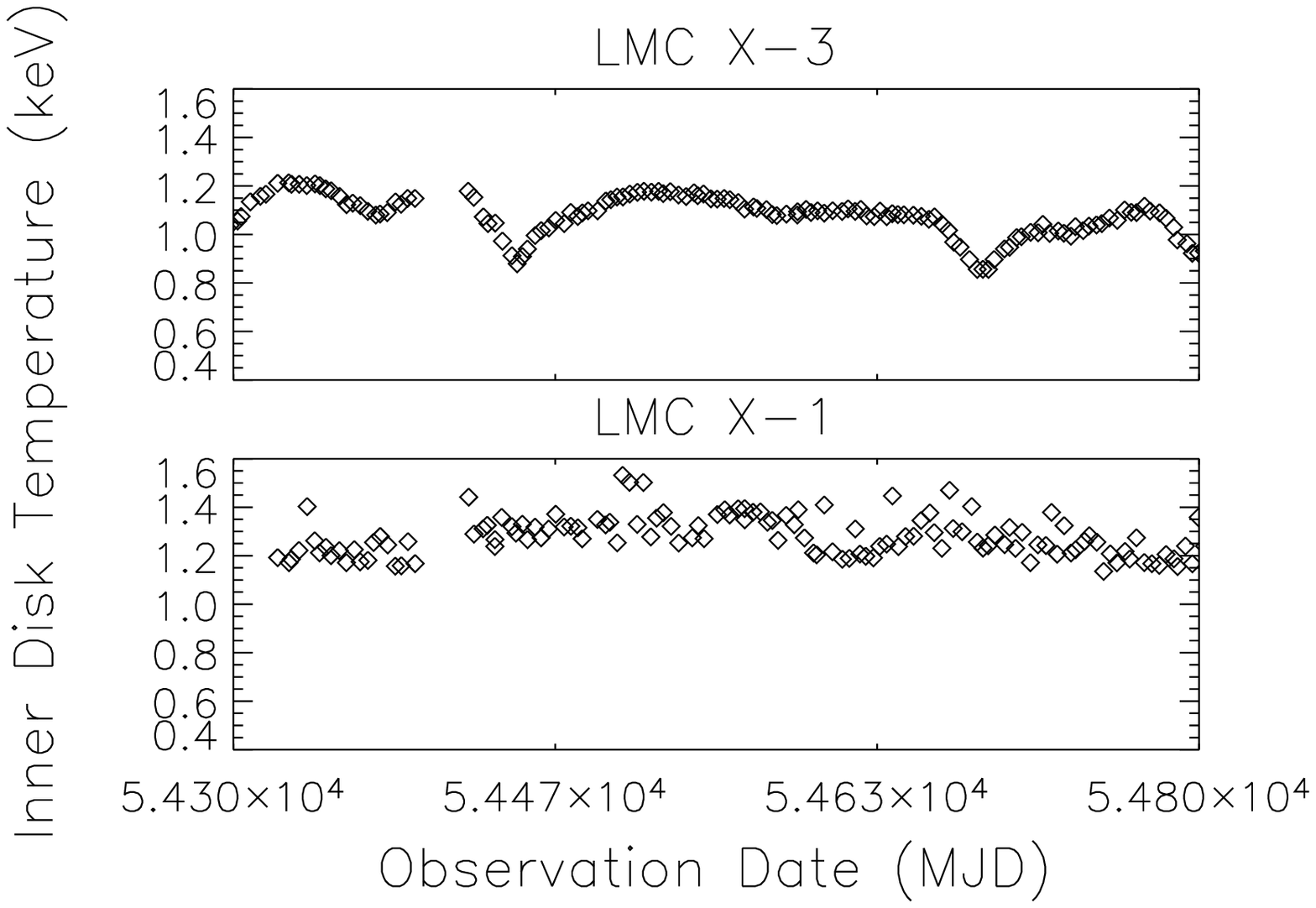}
  \caption{The innermost disk temperature plotted as a function of
  observation date for \protect\xthree\ (top) and \protect\xone
  (bottom). \label{fig:lmcx1-and-lmcx3-disk-temp-vs-date-short}}   
\end{figure}

Figure \ref{fig:lmcx1-and-lmcx3-disk-temp-vs-date-short} compares how
the inner disk temperatures of \xthree\ and \xone\ vary over time. The
figure shows only a subset of the total observing campaign in order to
better display short term variations from each source.  The
change in disk temperature from one observation to the next is
markedly more continuous for \xthree\ than for LMC
X-1. Over longer time periods, however, \xthree's inner disk
temperature evolves significantly.  Such long-term disk temperature
evolution is absent in \xone.

\begin{figure}
  \includegraphics[width=0.7\textwidth]{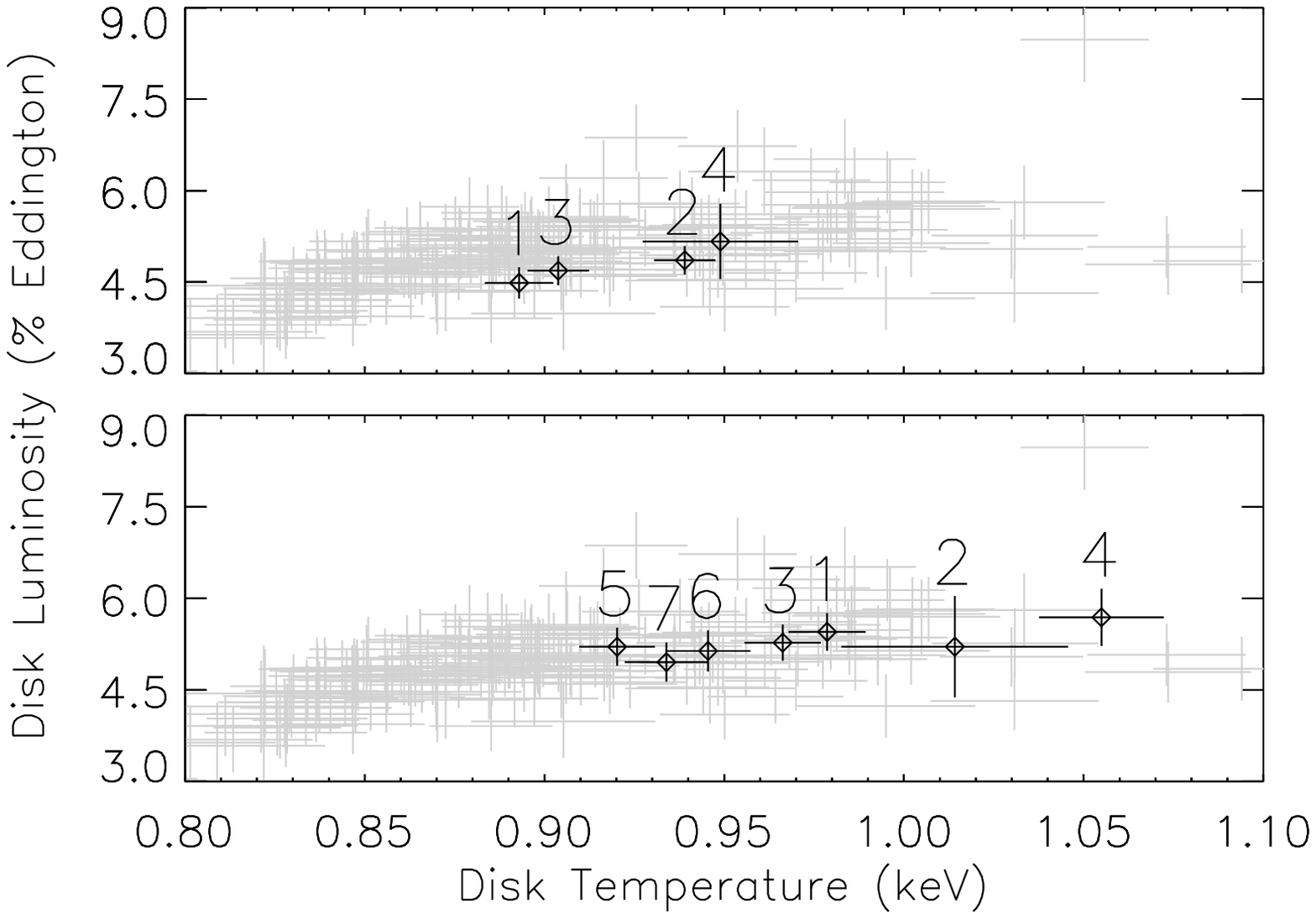}
  \caption{\protect\xone\ inner disk temperature vs. luminosity over the
  course of two representative archival observations.  The top panel shows
  observation 20188-01-05-00, while the lower panel shows observation
  30087-01-04-00.
  Black points,
  representing subsections of the archival observation, are numbered
   in chronological order. Gray points represent the data from our
  own observing campaign, and provide a basis for
  comparison.
  \label{fig:lmcx1-long-chopped-obs-with-our-data-as-background}} 
  \end{figure}

Figure \ref{fig:lmcx1-long-chopped-obs-with-our-data-as-background}
shows how \xone's inner disk temperature and 
luminosity varied over the course of two of the archival
observations listed in Table \ref{tb:archival-observations}. The
bottom panel shows data from the longest observation, while
the top panel shows data from an observation with a typical number of
exposure intervals.

The inner disk appears to evolve over a
period of several hours. Rather than evolving along the
Stefan-Boltzmann relation, the disk temperature and luminosity
fall along a line of a very different slope. The order of the points
is not uniform along this line either, suggesting motion faster than
has been resolved by the data.

\begin{figure}
  \includegraphics[width=1.0\textwidth]{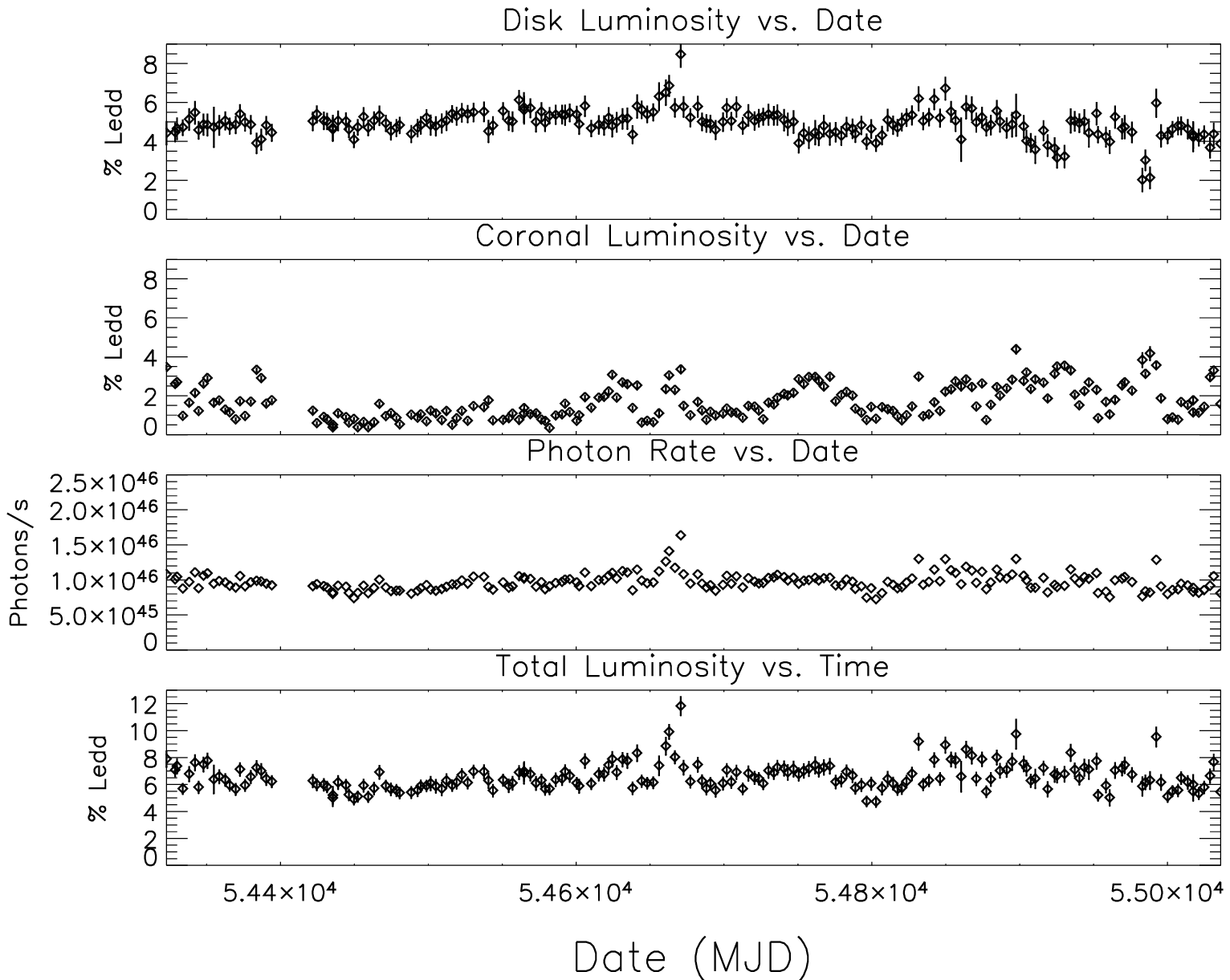}
  \caption{From top to bottom: disk luminosity, coronal luminosity,
  total photon emission rate, and total luminosity of \protect\xone
  vs. observation date.  Luminosities are shown in terms of the
  source's Eddington luminosity. \label{fig:lmcx1-4-panel-plot}}
\end{figure}

Figure \ref{fig:lmcx1-4-panel-plot} shows \xone's 
variability in disk, coronal, and total luminosities, as well as in
the total photon emission rate, over the course of our observing
program.  \response{The mild discrepancy between LMC X-1's overall accretion
  luminosity as presented
  here and in \citet{lmcx1-gou-2009} comes from the 
  narrower energy range we employ in calculating the source's overall
  flux. Both analyses, 
  however, find similar disk-to-coronal flux ratios in LMC X-1.} 
Percent variation, defined as
$\textrm{standard deviation} / \textrm{mean}$, quantifies the extent
to which the values in each panel vary.  The system's disk luminosity
shows $14.5$ \% variation, while its coronal up-scattering luminosity
shows $51.8$ \% variation.  
The system's overall x-ray luminosity has $15.2$ \%
variation. The total photon flux from LMC~X-1 is the system's most
stable feature, with only $12.5$ \% variation. While the total number
of photons emanating from the system   
remains relatively steady, the amount of energy that the corona
contributes to the total emission varies significantly.

\begin{figure}
  \includegraphics[width=0.7\textwidth]{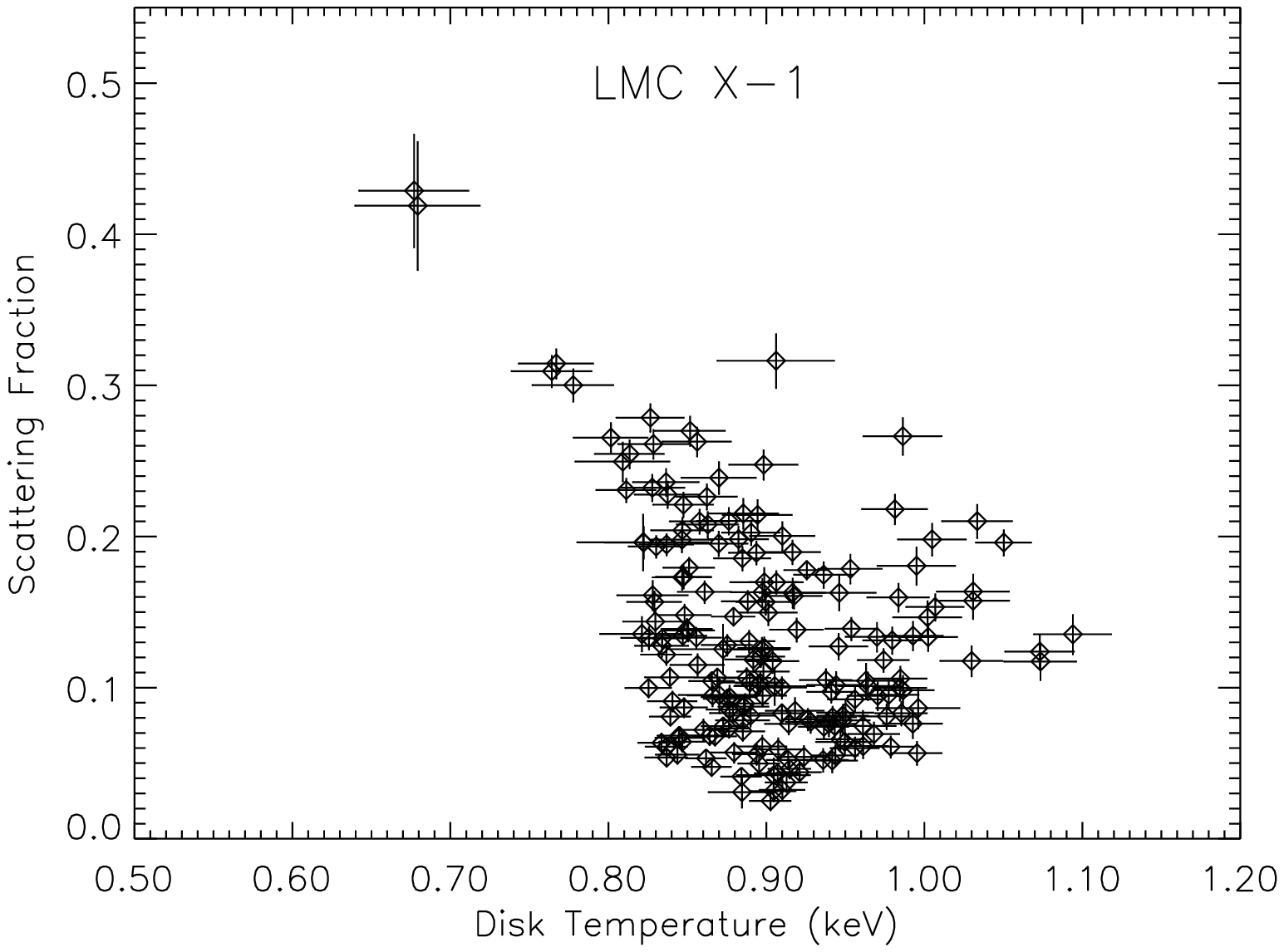}
  \caption{The fraction of total disk seed photons that undergo
  Compton up-scattering versus the inner disk temperature.   
    \label{fig:lmcx1-scattering-frac-vs-disk-temp}}
\end{figure}

Figure \ref{fig:lmcx1-scattering-frac-vs-disk-temp} displays the
relation between \xone's inner disk temperature and the 
inverse Compton scattering fraction. 
While observations with cooler disk temperatures and higher
scattering fractions suffer larger fitting uncertainties, 
\response{a Pearson correlation analysis confirms the existance of a
  statistically significant negative correlation between these two
  parameters, even when the two upper-leftmost points in Figure
  \ref{fig:lmcx1-scattering-frac-vs-disk-temp} are removed.}
Since the scattering fraction maps directly to the inverse Compton optical
depth, Figure \ref{fig:lmcx1-scattering-frac-vs-disk-temp} shows
that the inner edge of the disk appears cooler when there is more
coronal material.

\section{Discussion and Conclusions}

LMC~X--1's apparent defiance of the Stefan-Boltzmann relation
\response{may 
stem} from variations in the disk itself, or from variations in the
scattering medium surrounding it.  Because the total number
of photons 
is more nearly constant than the disk luminosity (see Figure
\ref{fig:lmcx1-4-panel-plot}), variations in the number of photons
diverted to the power law by the corona seem more likely to explain
the anomalous temperature/luminosity relations than variations within
the disk itself.

The first step in quantitatively distinguishing between these two
explanations
for the anomalous disk relation is to compare the observed disk
variation timescales to timescales inherent to the disk and to the
corona.  In Figure
\ref{fig:lmcx1-long-chopped-obs-with-our-data-as-background}, \xone\ 's
disk randomly samples a wide swath of the temperature
parameter space in the space of a few hours. Individual points in
the archival observation are separated by an average of 2320~s,
which provides an estimate of the timescale
for apparent disk variations.  We can compare this value to the disk's
viscous timescale to assess whether the disk's structure is
capable of changing over such a brief interval of time.

The disk's viscous timescale scales with both radius and viscosity, as \citep{frank-king-and-raine} 
\begin{equation}
t_{visc} \sim \frac{R^2}{\nu} 
\label{eq:viscous-timescale}
\end{equation}
The circularization radius, $R_{circ}$, of the accreted wind material
gives a lower limit for the outer disk radius.  The disk
radius, and therefore the viscous timescale, may actually be larger
due to outward angular momentum transfer.  Since we are primarily
concerned with finding a lower limit on the viscous timescale, however,
 $R_{circ}$ provides a conservative value of the disk radius. The
circularization radius is defined as the radius at which the accreted
matter's angular momentum due to Keplerian motion equals the angular
momentum it carried upon initial capture by the black hole.
Following \citet{frank-king-and-raine} we assume that the wind's
initial angular momentum about the black hole goes as 
\begin{equation}
  l \sim \frac{1}{4}\omega R_{acc}^2
  \label{eq:orbital_ang_mom}
\end{equation}
where $R_{acc}$ represents the black hole's gravitational capture
radius, and $\omega$ is the black hole's orbital angular velocity
about the companion.
From there, we obtain
\begin{equation}
  R_{circ} = \frac{G^3 M_{BH}^3 \omega^2}{v_{rel}^8} \textrm{.}
  \label{eq:Rcirc}
\end{equation}
where $M_{BH}$ represents mass of the black hole, and 
\begin{equation}
  v_{rel}^2 = v_w^2 + v_{orb}^2 \textrm{.}
  \label{eq:vrel}
\end{equation}
Therefore, the disk's size is strongly dependent on the speed of the
stellar wind, which undergoes line-driven acceleration
\citep{CAK, winds-owocki-1994, winds-kudritzki-2000}. 
The wind's velocity increases with distance from the star
\citep{lamers-1999, kudritzki-1989} as
\begin{equation}
  v_w(r) = v_{\infty} \left[1 - 0.9983 \frac{R_*}{r} \right]^{\beta} \textrm{.}
  \label{eq:vwind}
\end{equation}
Here, $ v_{\infty}$ is the wind's terminal velocity and $R_*$ is the
companion's radius.  The factor of $0.9983$ is chosen such that
$v_{w}$ equals the escape velocity at the star's surface.  The values
of $v_{\infty}$ and $\beta$ are poorly constrained for O-star winds,
but typically range from 1000--2000 km/s and from 0.5--1.5,
respectively \citep{kudritzki-1989},\citep{lamers-1999}.  We conservatively assume $v_{\infty} = 1700$ km/s,
in order to compare our results with other works that consider
clumping in O-star winds \citep{the-italians}. We carry the effects of
the full range of $\beta$ 
values through all subsequent calculations to obtain a range of results.

The final step in calculating $R_{circ}$ is determining $v_w$ where the
wind reaches the accretion cylinder.
To find $R_{acc}$, we equate the wind's
gravitational and kinetic energy terms with respect to the black hole
and find 
\begin{equation}
  R_{acc} = \frac{2GM_{BH}}{v_{rel}^2} \textrm{.}
  \label{eq:Racc}
\end{equation}
Solving (\ref{eq:vrel}), (\ref{eq:vwind}) and (\ref{eq:Racc}) numerically over the full
range of $\beta$ values constrains $v_w$ to lie
between $154$ and $635$ km/s by the time it reaches the black hole's
accretion cylinder.   $R_{circ}$ is thus limited to values
between $6.7 \times 10^8$ and $2.4 \times 10^{10}$ cm, in
agreement with previous work \citep{lmcx1-and-lmcx3-nowak-2000}.

Having found $R_{acc}$, we turn to calculating the disk's
viscosity, $\nu$.  The Shakura-Sunyaev $\alpha$-disk prescription
posits 
\begin{equation}
  \nu = \alpha c_s H
  \label{eq:alpha-prescription}
\end{equation}
where $c_s$ and $H$ are the disk's local sound speed and height,
respectively. Observations of transient x-ray binary outbursts
indicate $\alpha \sim 0.1$ \citep{alpha-disk-values}. Like \xone,
x-ray transients while in outburst host ionized accretion disks around
a central black hole.
These similarities motivate us to use $\alpha = 0.1$ in the following
calculations. 
Substituting equilibrium thin-disk formulae \citep{frank-king-and-raine}
for the sound speed and
disk height into (\ref{eq:alpha-prescription}) gives
\begin{equation}
  \nu = 1.8 \times 10^{14} \alpha^{4/5}\dot{M}_{16}^{3/10} m_1^{-1/4}
  R_{10}^{3/4} \left[1 - \left(\frac{R_{min}}{R} \right)^{1/2}
  \right]^{3/10} \textrm{ cm}^2\textrm{/g} \textrm{.} 
  \label{eq:alpha-disk-viscosity-formula}
\end{equation}
Here, $m_1$ is $M_{BH}$ in units of $M_{\sun}$, $\dot{M}_{16}$ is the
accretion rate in units of $10^{16}$ grams per second, $R_{10}$ is radius in
units of $10^{10}$ cm, and $R_{min}$ is radius of the innermost stable
orbit. Simple geometry requires that the accretion rate depends
on the companion's mass loss rate as
\begin{equation}
      \dot{M}_{BH} = \frac{1}{4} \left(\frac{R_{acc}}{R_{sys}}\right)^2
      \dot{M}_{*} \textrm{ g/s} 
  \label{eq:MdotBH}
\end{equation}
with a typical O-star $\dot{M}_*$ of $10^{-5}$ $M_{\sun}$ per year. 
The factor of
four in the denominator reflects the the fact that the black hole
presents a circular cross-section to the spherically expanding wind.
Equation (\ref{eq:MdotBH}) shows that the accretion rate onto the
black hole ranges between $2.2 \times 10^{17}$ and $9.6 \times
10^{18}$ g/s,
depending on the chosen value of $\beta$.  These values are in good
agreement with accretion rates   
derived from the observed x-ray luminosity using
\begin{equation}
  L \sim 0.1 c^2 \dot{M}_{BH},
  \label{eq:lum}
\end{equation} 
which range between $8.4 \times 10^{17}$ and $2.0 \times 10^{18}$ g/s. The $0.1$ 
efficiency factor in (\ref{eq:lum}) is recommended by \citet{frank-king-and-raine}. 
Equation \ref{eq:lum} arises from the condition that all gravitational
potential energy up to the last stable orbit is radiated away.

Using (\ref{eq:viscous-timescale}) to combine the results of
(\ref{eq:alpha-disk-viscosity-formula}) and (\ref{eq:Rcirc}) gives
$5\times 10^4 \le t_{visc} \le 4\times 10^6$ seconds.  These values of
$t_{visc}$ are at least an order of magnitude longer than the
$\sim 10^3$ second timescale of inner disk variation indicated by Figure
\ref{fig:lmcx1-long-chopped-obs-with-our-data-as-background}. This
provides the first piece of evidence that the observed disk
luminosity variations are not caused by a process inherent to the
disk itself.
But the calculation of the disk viscous
timescale above assumes that the entire disk is involved in any change.
The Roche-lobe accreting black-hole binary GRS~1915+105 shows very 
fast spectral changes in many of its variability
states near the Eddington luminosity \citep{grs1915-belloni-2000}, which
appear to be repeated ejections of the innermost parts of the disk.  But
when the inner disk is ejected and the spectrum turns hard in GRS~1915+105,
the total x-ray count rate also drops dramatically.  It is the absence
of any such change in count rate in \xone\ that leads us to favor obscuration
by marginally optically thick, ionized coronal material to an ejection
mechanism.

The slope of the temperature-luminosity relation that the chopped
archival observations in 
Figure \ref{fig:lmcx1-long-chopped-obs-with-our-data-as-background} occupy
provides the second indication that \xone's apparent disk luminosity
variations are caused by a process unrelated to the accretion disk.
If the variations in disk
luminosity were due to changes in the accretion rate, the points
would move sequentially along the Stefan-Boltzmann relation shown in 
Figures \ref{fig:lmcx3-disktemp-vs-lum-for-flash} and
\ref{fig:lmcx1-disktemp-vs-lum-for-flash}.

Since the anomalous disk temperature-luminosity relation is poorly
explained by changes within the accretion disk, we ask whether changes
in the corona could explain the relation.  First, we consider which
coronal properties are compatible with the observations.  Then, we
speculate how such a corona might form.

Figure \ref{fig:lmcx1-scattering-frac-vs-disk-temp} provides a clue as
to how the coronal material factors in \xone's anomalous disk 
temperature-luminosity relation. The fraction of disk photons that
suffer inverse Compton scattering while traversing the corona is
lowest when the apparent inner disk temperature is around $0.9$ keV.
This value lies at the higher end of the observed range of disk
temperatures for this source.  The inverse Compton scattering fraction
increases with decreasing inner disk temperature, up to a scattering
fraction of nearly unity at $T_{in}$ of $\sim 0.4$ keV. Such behavior
is consistent with a steady disk whose inner radii are sometimes
obscured by a cloud of energetic electrons.  In that case,
the number of disk photons, which form the seed photons for
inverse Compton scattering in the overlying corona, would remain
constant. The fraction of original photons that suffer up-scattering
would depend only on the amount of coronal material present.   
Figure \ref{fig:lmcx1-4-panel-plot} shows that \xone's total photon flux
remains largely constant, while its coronal luminosity varies
substantially.  

\begin{figure}
  \includegraphics[width=0.9\textwidth]{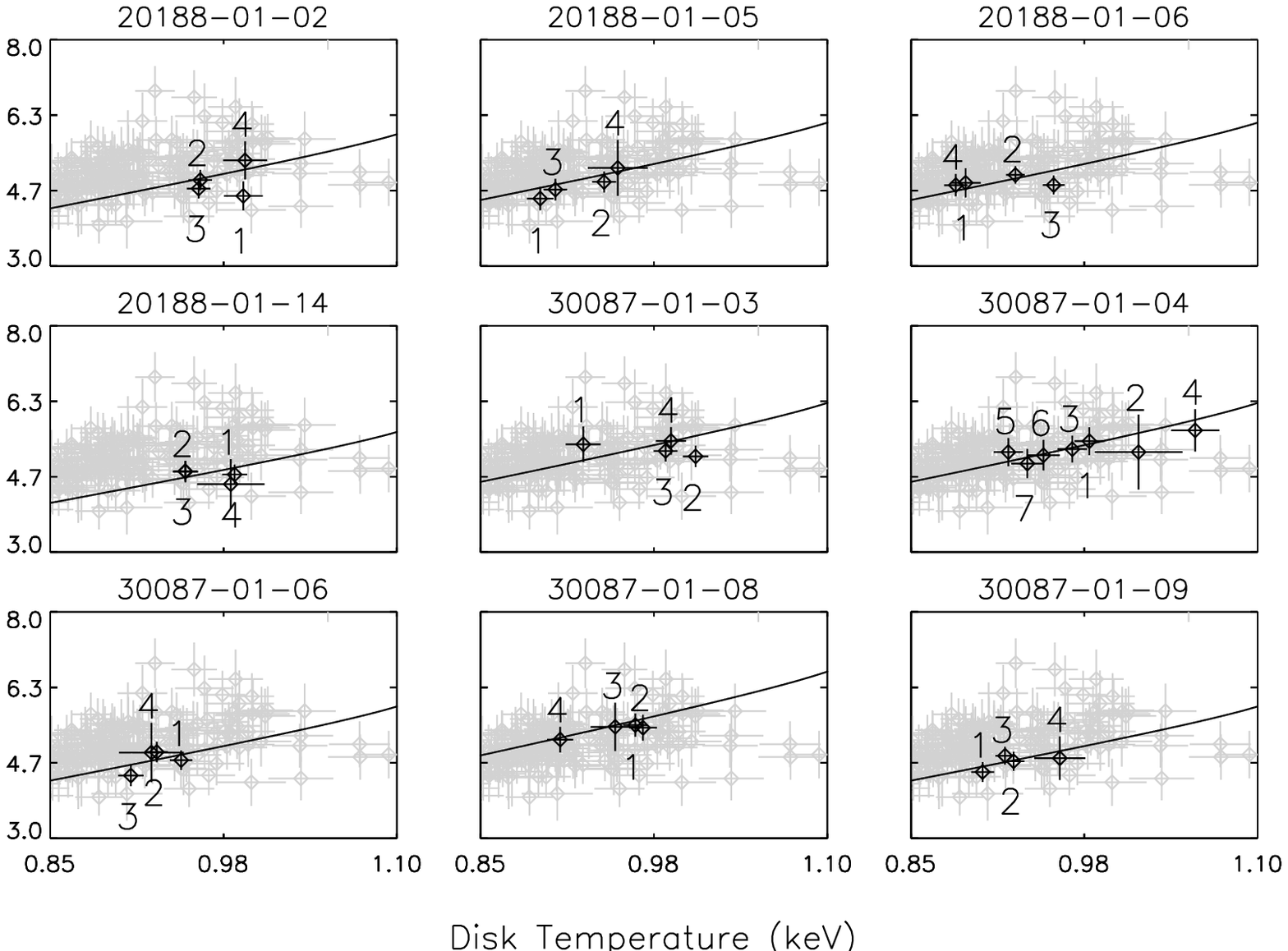}
  \caption{The obscured inner disk model, fitted to all of the chopped
    archival observations containing more than three sub-intervals.
    Observation numbers are listed in the upper left corner of each
    plot.  Black points represent the chopped observations, while
    gray points represent our data.  The solid black line plots the
    obscured inner disk model, which has been normalized to fit
    the chopped observations. 
    \label{fig:lmcx1-disk-lum-frac-change}} 
\end{figure}

To quantify this model, we integrate the emission from the 
$\alpha-$disk flux prescription \citep{frank-king-and-raine}
 outwards from inner radii at temperatures
ranging between $0.4$ and $0.9$ keV.  Figure
\ref{fig:lmcx1-disk-lum-frac-change} shows the resulting
temperature-luminosity relation, as fitted to all of the chopped archival
datasets with four or more sub-intervals.  
The disk temperature and luminosity are constant 
in each observation.  The amount of inner disk obscuration is the only
variable that determines where the chopped observations fall along the
best-fit line.   The model appears consistent with all the data, 
particularly with observation 30087-01-04, which provides the strongest
constraint.  Note that these individual chopped observations run
roughly perpendicular to the Stefan-Boltzmann relation in
temperature-luminosity space.  Suzaku or NuSTAR
observations may allow the detection of variations of the power-law
index on relevant (hour) timescales; hardness variations might then be
interpreted as variations in Comptonizing optical depth, in which case
hardness might be expected to correlate with the flux in the tail.

Figures \ref{fig:lmcx1-disk-lum-frac-change} and
\ref{fig:lmcx1-scattering-frac-vs-disk-temp} raise questions about the 
analysis in \citet{lmcx1-gou-2009}, which derives the spin of LMC~X--1's
black hole from measurements of the inner disk temperature. 
Their analysis requires that the inner disk always be unobscured. Our
analysis, however, shows that the degree of inner disk obscuration in
LMC~X--1 varies significantly on short timescales.  In systems such as
LMC~X--3, where the disk is not obscured, there is no difficulty with
the approach of \citet{lmcx1-gou-2009}.

Why, in fact, does LMC~X--3 have less apparent inner disk obscuration
than LMC~X--1 does?   We can imagine \response{at least} three
possible answers. The first option 
is that the coronal optical depth is much larger, in general, in
LMC~X--1 than in LMC~X--3.  This possibility is ruled out by the
overlap between the two sources in terms of the flux ratio between the
disk and power-law components;  when the disk luminosity in \xthree
is in decline, a significant power-law tail can form in that system
without disrupting the excellent Stefan-Boltzmann relation for the
disk component \citep{smith-2007}.
A second possible answer supposes that the geometries of the coronae
are very different: LMC~X-3 has a geometrically large corona, while
LMC~X--1 has one that is more centrally concentrated, so that it
obscures more of the inner disk while producing the same amount of 
upscattering.
A third possibility supposes that the corona in both cases has a
conelike shape (possibly a jet, but not necessarily, from our
evidence alone), and that the apparent differences between the two sources
are due to an inclination effect.  In \xone, which has an inclination
of $36.38\pm 1.92^{\circ}$ \citep{lmcx1-orosz-2009}, the conical
corona may block the inner parts of the disk as we look down through
it, while in
LMC~X--3, with an inclination between $50^{\circ}$ and
$70^{\circ}$ \citep{lmcx3-vanderKlis-1985}, the cone presents
itself in semi-profile, where it doesn't obscure the inner disk but
we can still see the Compton upscattered flux it produces coming out
its sides.  

While either of the latter two pictures is consistent
with our data, the last is more appealing \response{for two
  reasons. Firstly}, it 
does not require 
an unexplained difference in the form of the corona in the two
binaries.  
\response{Secondly, it may someday be possible to verify or to disprove
  using x-ray polarization observations. A 'down the barrel' 
  jet observation (e.g. LMC~X--1) would sport a lower over-all scattering
  polarization fraction than an observation of the same type of
  geometry viewed 'in profile' (e.g. LMC~X--3) would.  
  The precise magnitude of the expected polarization fraction
  difference will govern how easily its presence or absence can be
  verified. 
  In any case, future observations
  with the GEMS x-ray polarimeter will at
  least begin to explore the relevant parameter space, provided 
  the instrument can spend upwards of $10^5$ seconds observing each
  source \citep{GEMS-paper}.}

\begin{figure}
  \includegraphics[width=0.7\textwidth]{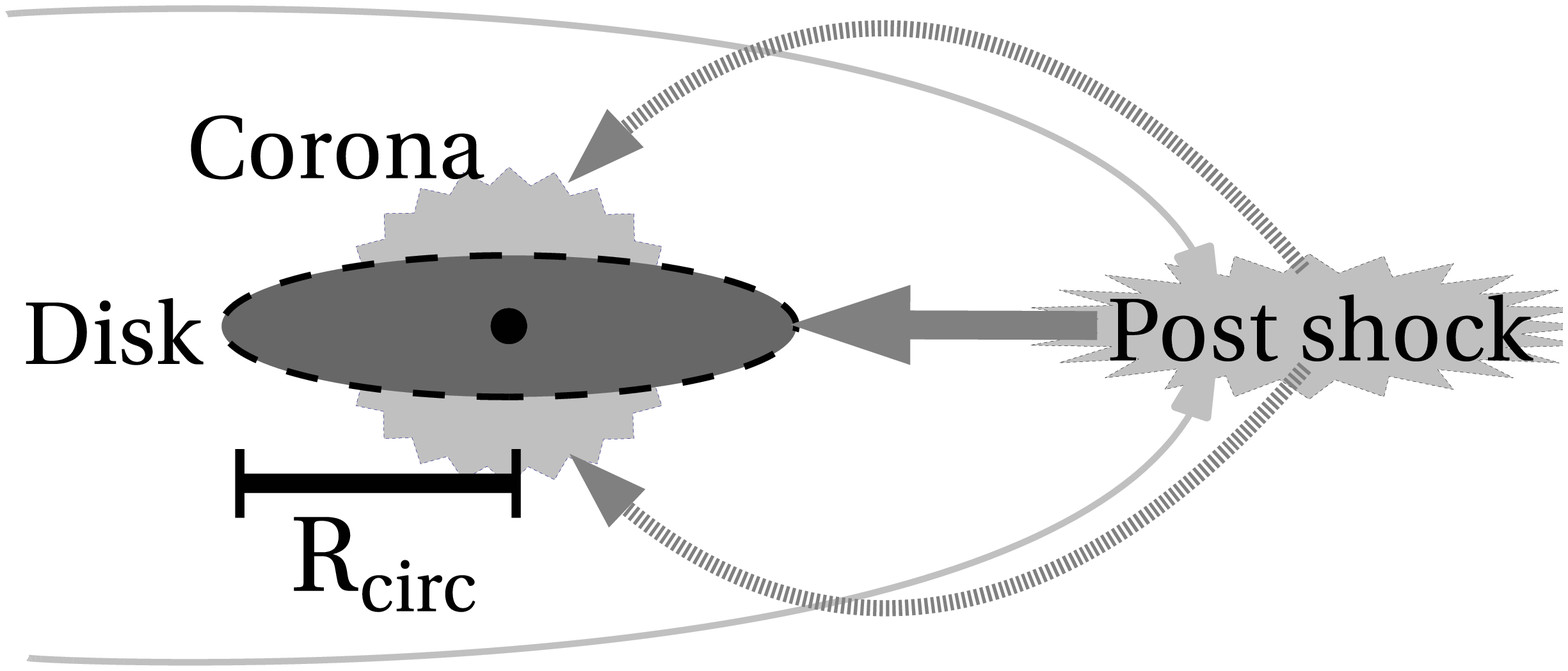}
  \caption{A cartoon of the accretion geometry of wind
  accretion systems.  Incoming material from the companion's wind
  (dashed gray lines) must
  dissipate its angular momentum in a shock front behind the compact
  object, a.k.a. the 'post shock',  before it can be accreted (thick
  arrow). \label{fig:wind-accretion-geometry-cartoon}} 
\end{figure}

Finally, we turn to the question of how a population of energetic
electrons forms and dissipates above the innermost region of the
disk on timescales shorter than $t_{visc}$.  
Figure \ref{fig:wind-accretion-geometry-cartoon} illustrates one
possible accretion scenario that could account for LMC~X--1's unusual
properties.  

In this picture, the incoming wind material sheds its net angular
momentum in the ``post shock'' before accreting onto the black
hole \citep{hoyle-lyttleton-accretion}. Most of the shocked gas
remains in the disk plane, and enters 
the accretion disk (middle arrow). A smaller portion of the heated  
post-shock matter launches on randomly distributed radial 
trajectories towards the black hole (upper and lower arrows).  This
component of the accretion flow rapidly heats on its radial descent
into the black hole's gravitational potential well.  Unlike the disk,
which is dense enough to cool efficiently via bremsstrahlung, the
diffuse infalling matter cannot efficiently cool via free-free
interactions.  By the time the radially accreting matter
passes over the inner disk, it can have reached electron temperatures above
the $100$ keV required for the observed inverse Compton upscattering
\citep{narayan-and-yi-1995,ichimaru-1977,rees-1982,esin-1998}.   
This model allows the corona to change independently of
the disk.

The presence of two accretion flow
components, which are largely independent beyond the
post-shock, is consistent with the constant observed total photon
number and minimal disk luminosity variation in \xone. Changes in the 
wind mass flux, due perhaps to wind clumping, register
immediately in the radial, coronal flow.  Viscous diffusion in the
disk, meanwhile, smooths out those same sharp mass discontinuities.
While the coronal surface density, and therefore the total energy of
coronal emission changes rapidly, the disk emission evolves more
slowly and less dramatically. The scenario of independent flows with
the thin disk remaining intact beneath the corona, even when the
latter is optically thick, is consistent with the picture derived from
observations of the hard state and hard-to-soft transitions in 1E
1740.7-2942 and GRS 1758-258 \citep{smith-2002}.
Recent observations of the iron fluorescence line in black holes in
the hard state have shown relativistic line profiles indicating that
the inner disk remains present even in the hard state
\citep{miller06}, until extremely low lumniosities, below 1\% of
Eddington, are reached \citep{tomsick09}.  In the luminosity range of
LMC~X--1, we therefore expect the inner disk to remain present,
whether the thin disk is actually continuous or whether its inner
portion is recondensed from a disk that is disrupted at intermediate
radii \citep{liu11}.

Differing patterns of x-ray variability 
(e.g. Figure~\ref{fig:lmcx1-and-lmcx3-disk-temp-vs-date-short}) 
show promise as a means of identifying black hole companions when optical
observations cannot.
Fast changes in the inner disk temperature
(Figure~\ref{fig:lmcx1-and-lmcx3-disk-temp-vs-date-short}), and an
anomalous temperature/luminosity relation
(Figure~\ref{fig:lmcx1-disktemp-vs-lum-for-flash}),
both combined with a stable net photon flux
(Figure~\ref{fig:lmcx1-4-panel-plot}), may signal the presence of a
wind accretor, perhaps combined with a low inclination, just as
long-term hysteresis in state changes appears to be a signature of
Roche-lobe overflow accretion and a large disk \citep{smith-2002,
smith-2007}.

The type of two-component accretion flow proposed in this paper is
best suited to black hole 
binary systems where the companion drives a high-velocity stellar wind.
At the same time, this particular model leaves ample room for other
coronal generation mechanisms that have already been proposed for 
black hole binary systems accreting via
Roche-lobe overflow.  

At least two types of  observations can either help to confirm
or to disprove the hypothesis we have proposed. 
The first focuses on recent x-ray observations of
IC~10~X--1 and NGC~300~X--1 \citep{ic10x1-barnard-2008}. Both systems
have disk to power-law luminosity ratios, as well as high-mass
main-sequence companions, that resemble \xone's.
Binary systems that contain both an O-type
star capable of driving such a wind and a black hole companion are
rare, which has made this particular combination of accretion mechanisms
particularly hard to observe. Further observations of IC~10~X--1 and
NGC~300~X--1, however, can potentially determine 
how deep their apparent similarities to \xone\ run, and will provide
more opportunities to test our model's veracity.
The second test relies on observations of intermediate
state, low-disk-fraction black hole binaries. These systems exhibit
unusual temperature-luminosity relations akin to the ones seen in
\xone\ \citep{dunn-2011}. One can confirm whether the same mechanism  
is at work in both \xone\ and an intermediate state system by measuring the  
variability (or lack thereof) of the latter's total photon flux.  Stable 
photon fluxes would support the proposition that these systems also
contain stable, but variably-occulted, inner disks.

\xone\ has provided specific evidence that stellar mass black
hole binaries may have x-ray spectral features which are unique
to wind accretion.
In a more general sense, the fact that \xone\ shares certain x-ray
spectral properties with other types of black hole binary systems
motivates further investigation of its seemingly unorthodox accretion
mechanism's full scope and utility.

\begin{acknowledgements}
This work was supported by NASA grant NNX09AC86G. The authors thank
the anonymous referree for his or her constructive and insightful
suggestions. 
L.R. thanks
E. Ramirez-Ruiz, R. Strickler, and J. Naiman for productive discussions.    
\end{acknowledgements}

\bibliographystyle{apj}
\bibliography{LMCX1-Ruhlen-Smith-revised}
\end{singlespace}
\end{document}